\documentclass[fleqn,usenatbib]{mnras}
\usepackage{newtxtext,newtxmath}
\usepackage[T1]{fontenc}

\DeclareRobustCommand{\VAN}[3]{#2}
\let\VANthebibliography\thebibliography
\def\thebibliography{\DeclareRobustCommand{\VAN}[3]{##3}\VANthebibliography}

\usepackage{indentfirst}
\usepackage{physics}
\usepackage{ulem}
\usepackage{graphicx}	% Including figure files
\usepackage{amsmath}	% Advanced maths commands
\usepackage{xcolor}

\title[Metacalibration for the Roman HLIS]{Weak gravitational lensing shear estimation with \textsc{metacalibration} for the \emph{Roman} High-Latitude Imaging Survey}

\author[M. Yamamoto et al.]{
Masaya Yamamoto,$^{1}$\thanks{E-mail: masaya.yamamoto@duke.edu}
M. A. Troxel,$^{1}$
Mike Jarvis,$^{2}$
Rachel Mandelbaum,$^{3}$
Christopher Hirata,$^{4,5,6}$
\newauthor{
Heyang Long,$^{4,5}$ Ami Choi,$^{7}$
and Tianqing Zhang$^{3}$}\\
$^{1}$Department of Physics, Duke University, Durham, NC, 27708\\
$^{2}$Department of Physics and Astronomy, University of Pennsylvania, Philadelphia, PA 19104, USA\\
$^{3}$McWilliams Center for Cosmology, Department of Physics, Carnegie Mellon University, Pittsburgh, Pennsylvania 15213, USA\\
$^{4}$Center for Cosmology and Astro-Particle Physics, The Ohio State University, 191 West Woodruff Avenue, Columbus, OH 43210, USA\\
$^{5}$Department of Physics, The Ohio State University, 191 West Woodruff Avenue, Columbus, OH 43210, USA\\
$^{6}$Department of Astronomy, The Ohio State University, 140 West 18th Avenue, Columbus, OH 43210, USA\\
$^{7}$Department of Physics, California Institute of Technology, 1200 E. California Blvd., Pasadena, California 91125, USA\\
}

\date{Accepted XXX. Received YYY; in original form ZZZ}

\pubyear{2015}

\begin{document}
\label{firstpage}
\pagerange{\pageref{firstpage}--\pageref{lastpage}}
\maketitle

\begin{abstract}
We investigate the performance of the \textsc{metacalibration} shear calibration framework using simulated imaging data for the \emph{Nancy Grace Roman} Space Telescope (\emph{Roman}) reference High-Latitude Imaging Survey (HLIS). The weak lensing program of \emph{Roman} requires the mean weak lensing shear estimate to be calibrated within about 0.03\%. To reach this goal, we can test our calibration process with various simulations and ultimately isolate the sources of residual shear biases in order to improve our methods. In this work, we build on the HLIS image simulation pipeline to incorporate several more realistic processing-pipeline updates. We show the first \textsc{metacalibration} results for six \textrm{deg}$^2$ of the simulated reference HLIS and compare them to measurements on simpler, faster \emph{Roman}-like image simulations. We neglect the impact of blending of objects. We find in the simplified simulations, \textsc{metacalibration} can calibrate shapes to within $m=(-0.01\pm 0.10)$\%. When applied to the current most-realistic version of the simulations, the precision is much lower, with estimates of $m=(-0.76\pm 0.43)$\% for joint multi-band multi-epoch measurements and $m=(-1.13\pm 0.60)$\% for multi-band coadd measurements. These results are all consistent with zero within 1--2$\sigma$, indicating we are currently limited by our simulated survey volume. Further work on testing the shear calibration methodology is necessary at the precision of the \emph{Roman} requirements, in particular in the presence of blending. Current results demonstrate, however, that \textsc{metacalibration} can work on undersampled space-based \emph{Roman} imaging data at levels comparable to requirements of current weak lensing surveys.
\end{abstract}


%%%%%%%%%%%%%%%%%%%%%%%%%%%%%%%%%%%%%%%%%%%%%%%%%%

%%%%%%%%%%%%%%%%% BODY OF PAPER %%%%%%%%%%%%%%%%%%

\section{Introduction}

Since the formulation of the standard cosmological model, most observational evidence has been consistent with the flat $\Lambda$-CDM model in which the contents of the universe are mostly in the dark sector (\citealt{2020A&A...641A...8P, 2020MNRAS.498.2492G, 2020MNRAS.499.5527T, 2021A&A...645A.104A, 2022PhRvD.105b3520A, 2022arXiv220204077B}). Within this sector, dark energy accounts for the accelerating expansion of the universe (e.g., \citealt{1998AJ....116.1009R, 1999AIPC..478..129P}) and the combination of dark energy and dark matter are responsible for the observed growth of large-scale structure (\citealt{2015RPPh...78h6901K, 2017grle.book.....D}). The growth of structure can be studied by measuring quantities that are sensitive to the way light rays are deflected by intervening mass in the Universe, a physical phenomenon called gravitational lensing. “Weak” gravitational lensing causes a slight distortion to intrinsic galaxy shapes called shear. By detecting millions of these gravitationally lensed galaxies and computing their ensemble shear, we can explore the dark matter density and the amplitude of matter fluctuation in the universe (e.g., \citealt{2001PhR...340..291B}). Measuring these quantities eventually helps us learn about the history of the development of the large-scale structure in the universe. For that reason, weak lensing is one of the most powerful probes in current and near-future imaging surveys to constrain cosmological parameters with high precision.

Due to the subtlety in the lensing effect and its systematics-dominant nature in the weak regime (\citealt{2018ARA&A..56..393M}), observational efforts have been challenging. Over the past decade, large international collaborations such as the Dark Energy Survey\footnote{\url{http://www.darkenergysurvey.org/}} (DES: \citealt{2005astro.ph.10346T}), the Hyper Suprime-Cam Subaru Strategic Program\footnote{\url{http://hsc.mtk.nao.ac.jp/ssp/}} (HSC: \citealt{2018PASJ...70S...4A}), and the Kilo-Degree Survey\footnote{\url{http://kids.strw.leidenuniv.nl/}} (KiDS: \citealt{2013ExA....35...25D}) have been successful at constraining the cosmological parameters, and the precision in calibrating the estimates of shear from the shapes of distant galaxies has reached a few percent (\citealt{2018MNRAS.481.3170M, 2020arXiv201103408G, 2020arXiv201208567M, 2021A&A...645A.105G}). As we observe more area on the sky and develop better tools (observatories and algorithms), we detect more galaxies, and have already reached the point where statistical and systematic uncertainties are comparable. Thus, near-future Stage IV surveys (\citealt{2006astro.ph..9591A}) such as Euclid\footnote{\url{ http://sci.esa.int/euclid}} (\citealt{2011arXiv1110.3193L}), the Vera C. Rubin Observatory Legacy Survey of Space and Time\footnote{\url{ http://www.lsst.org}} (\emph{LSST}: \citealt{2009arXiv0912.0201L, 2019ApJ...873..111I}), and the \emph{Nancy Grace Roman} Space Telescope\footnote{\url{https://roman.gsfc.nasa.gov}} (\emph{Roman}: \citealt{2015arXiv150303757S}) require even better control of systematics. In order to understand the systematic uncertainties we face, we must develop realistic image simulations, apply existing shape measurement methods to the simulated images, and determine any potential residual systematic effects and build a strategy for modifying the calibration methodology to mitigate them.

Based on our current knowledge, systematic biases for weak lensing science, both observational and astrophysical, can occur at all stages of the imaging survey (e.g., \citealt{2018ARA&A..56..393M}). In particular, observational systematics can be due to: 
\begin{itemize}
    \item inhomogeneous observing strategy,
    \item atmospheric and instrumentation effects,
    \item post-processing pipelines such as image coaddition and object detection.
\end{itemize} 
Conventionally, these observational systematic biases are characterized through a number of simulations and validation tests, and these propagate into uncertainties on the final mean galaxy shear we measure. We quantify this impact as shear calibration bias. Several shear calibration and bias mitigation efforts have been inspired by the outcomes of the Shear TEsting Programme (STEP; \citealt{2006MNRAS.368.1323H, 2007MNRAS.376...13M}), the GRavitational lEnsing Accuracy Testing challenges (GREAT; \citealt{2010MNRAS.405.2044B, 2013ApJS..205...12K, 2015MNRAS.450.2963M}) and other image simulations work (e.g., \citealt{2017MNRAS.468.3295H}), to meet the requirements for the current imaging surveys. In particular, one of the state-of-the-art self-calibration methods, \textsc{metacalibration} (\citealt{2017arXiv170202600H, 2017ApJ...841...24S}) has been shown to be able to substantially reduce the significance of shear bias. It has been confirmed that shear can be calibrated with \textsc{metacalibration} at the few-percent level in DES (\citealt{2018MNRAS.481.1149Z, 2020arXiv201103408G}), without accounting for galaxy blending and detection. Tackling the issue of blending/detection is one key step forward for the DES Y6 analysis and preparatory Rubin LSST work with the shear-dependent detection and calibration technique \textsc{metadetection}\footnote{\textsc{metadetection} reduces a significant amount of shear-dependent detection bias by detecting objects after shearing a small region of the sky to measure the shear-dependent detection selection bias. } (\citealt{2020ApJ...902..138S, 2021A&A...646A.124H}). While the major advantage of \textsc{metacalibration} or \textsc{metadetection} is that they can be directly applied to real galaxy images without needing to rely on an ensemble calibration from image simulations, limitations for future surveys are not well-known. 

One possible limitation might lie in the effect of undersampled images in space-based surveys like \emph{Euclid} and \emph{Roman}, which will operate at their respective diffraction limits. The Point Spread Function (PSF) needs to be interpolated and estimated from well-sampled images to accurately deconvolve and measure shapes with, because the \emph{Roman} PSF has a complex structure that cannot be captured in the original undersampled images. It is, therefore, necessary to build a robust strategy to reconstruct well-sampled images to estimate the PSF to allow unbiased shape measurement.
Recently, \cite{2021MNRAS.502.4048K} (hereafter K21) addressed this potential issue of the limitations of \textsc{metacalibration} on undersampled images in \emph{Euclid} image simulations. They found that for the \emph{Euclid} mission the shear estimate with \textsc{metacalibration} is biased by about 1$\%$. It is mentioned that their result could be extended to the \emph{Roman} mission due to similarities in the instruments and for \emph{Roman} they predicted the multiplicative bias was more than 1\%. They show that these effects can be mitigated using additional weighting kernels in the measurement. 

In this work, we explore how \textsc{metacalibration} performs using coadd images at higher resolution than the native resolution of the instrument, taking advantage of the dithering of images in the reference HLIS. We use an updated suite of image simulations specifically made for the \emph{Roman} reference HLIS mission. Our work is based on the image simulation suite for the \emph{Roman} Space Telescope developed by \cite{2021MNRAS.501.2044T} (hereafter T21), where we render star and galaxy images using \texttt{GalSim}\footnote{Version 2.3.1 was used. \url{ https://github.com/GalSim-developers/GalSim}} (\citealt{2015A&C....10..121R}). We describe several important updates to the simulation capabilities and realism following T21. We also implement for the first time \textsc{metacalibration} using the simulated imaging within the T21 simulation suite, so that we can start to explore if the shear calibration goals of the HLIS for \emph{Roman} ($m=~3.2\times10^{-4}$) (\citealt{2018arXiv180403628D}) can be achieved with current \textsc{metacalibration} implementations,\footnote{We limit the study to \textsc{metacalibration} for now, since we can extract unblended cutouts of objects in our simulations.} or will require substantial additional development. \par

This paper is organized as follows. We first introduce the formalism of shear bias and image sampling relevant to the space-based surveys in Sec.~\ref{sec:background}. We then briefly discuss the simulation suite details we used and the updates we implemented for this project in Sec.~\ref{sec:sims}. In Sec.~\ref{sec:methods}, we show how coadditions of single exposure postage stamps are produced and how shape measurements are completed using \textsc{metacalibration}. We present the galaxy catalog properties and calibration results for single-band and multi-band measurements in Sec.~\ref{sec:results}. Finally, in Sec.~\ref{sec:discussion}, we discuss how we will be able to constrain the shear bias better in terms of further updates in the image simulations and what we can conclude from this study. 

\section{Background}
\label{sec:background}
In this section, we provide a brief introduction to image sampling defined through the Nyquist-Shannon Sampling Theorem, an overview of shear calibration bias within the context of weak lensing, and the \textsc{metacalibration} formalism.

\subsection{Image Sampling}
Based on the Nyquist-Shannon Sampling Theorem, which states that in order to reconstruct an unbiased continuous band-limited function without a loss of information, the sufficient sample rate is twice the bandlimiting frequency per second. In the context of image sampling, the criterion that the sample pixel spacing needs to satisfy is $p < \frac{1}{2u_{\text{max}}}$ ($u_{\text{max}}$ is the maximum spatial bandlimiting frequency). Since the spatial bandlimiting frequency in an astronomical image is defined as 
\begin{equation}
    u_{max} = \frac{1}{\lambda_{\text{min}}N_{f}}
\end{equation}
the sampling factor for a space telescope is defined as 
\begin{equation}
    Q = \frac{\lambda_{\text{min}}N_{f}}{p}, 
    \label{eqn:sampling}
\end{equation}
where $\lambda_{\text{min}}$ is the shortest wavelength of the incident light in each filter, $N_{f}$ is the focal ratio of the telescope ($N_{f}=7.8$ for \emph{Roman}) and $p$ is the pixel spacing of the sensor ($p=10\,\mu m$ for \emph{Roman}) (\citealt{2013PASP..125.1496S}). An image with sampling factor $Q=2$ is considered Nyquist-sampled and an image with sampling factor $Q<2$ is undersampled. The image sampling for \emph{Roman} are $Q$=0.88, 1.08, 1.31 for the J129, H158, and F184 bandpasses respectively.

\subsection{Shear Calibration Bias}
In the observation of weak lensing, we can quantify the bias associated with the shear recovery processes such as PSF estimation and shape measurement in image simulations. We call this shear calibration bias and we compute the deviations of the measured gravitational lensing shear $\gamma$ from the input shear. Here we define the reduced shear $g \equiv \gamma/(1-\kappa)$, and it is safe to approximate $\gamma \approx g$ in the context of cosmic shear. In the limit of weak lensing ($\lvert\gamma\rvert\ll1$, $\lvert\kappa\rvert\ll1$) and random intrinsic galaxy shapes, the ensemble average of ellipticities $\langle e \rangle$ of galaxy shapes is directly related to the reduced shear field $g$, and the estimated shear can be written as a linear model with multiplicative ($m_{i}$) and additive bias ($c_{i}$) as described in the equation below (\citealt{2006MNRAS.368.1323H, 2006MNRAS.366..101H, 2007MNRAS.376...13M}) 
\begin{equation}
    g^{\textrm{obs}}_{i} = (1+m_{i})g^{\textrm{true}}_{i} + c_{i}, 
    \label{eqn:linear}
\end{equation}
where $i$=(1,2) are the two components representing an elliptical distortion, which can be derived from the major and minor axes, and $g^{\textrm{obs}}_{i}$ is the two-component observed reduced shear after calibrations and $g^{\textrm{true}}_{i}$ is the true shear. The bias can be introduced in places such as PSF modeling, blending, and the undersampling of the image (e.g., \citealt{2018ARA&A..56..393M}). Additionally, complex detector effects such as the brighter-fatter effect are potential sources of bias (\citealt{2020PASP..132a4502C}), while it has been shown that for \emph{Roman} the effect of persistence will not be an issue (\citealt{2021arXiv210610273L}). Qualitatively, if the recovered shear is biased by 1$\%$ ($m=0.01$), from the cosmic shear power spectrum $S_{8} = \sigma_{8} \sqrt{\Omega_{m}/0.3}$ could be estimated to be biased about 1.5$\%$ in the final cosmology result. Thus, quantifying and correcting these biases before the real survey begins are extremely important. 

\begin{figure}
	\includegraphics[width=\columnwidth]{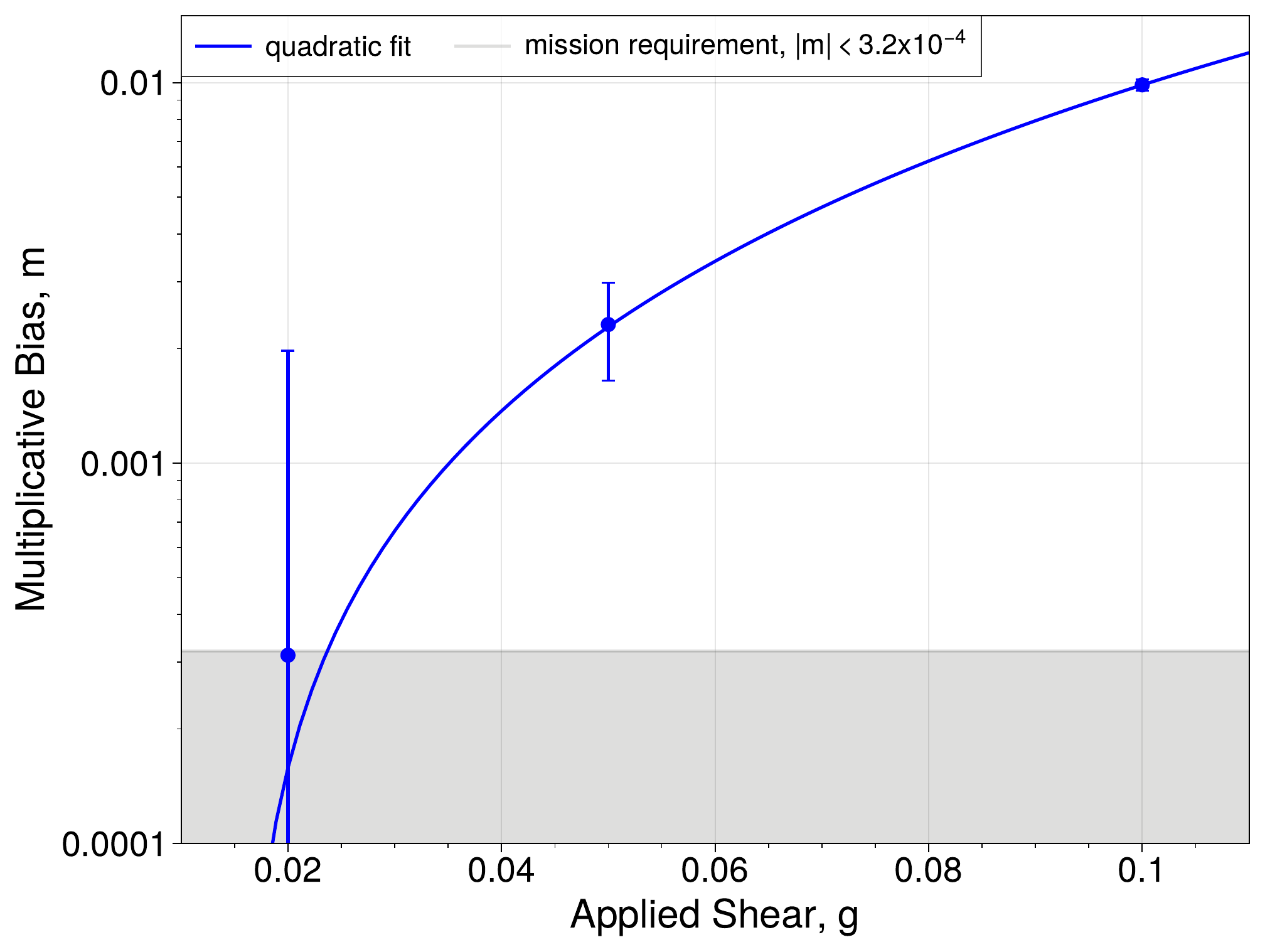}
    \caption{We show the recovered multiplicative bias on the shear as a function of the applied shear in our simplified simulations. The presented multiplicative bias is the average of $m_{1}$ and $m_{2}$, and the grey shaded region represents the mission requirement. At first order, the \textsc{metacalibration} algorithm only works for small input shears. The next order effect is seen as a quadratic dependence of the multiplicative shear bias on on the applied shear (\citealt{2017ApJ...841...24S}). A larger input shear increases the signal-to-noise of bias estimates, but to keep the non-linear effect small compared to requirements, we limit input shears to $\pm 0.02$. }
    \label{fig:metacal_shear_linear}
\end{figure}

\subsection{\textsc{metacalibration} Formalism}
Here, we briefly summarize how \textsc{metacalibration} (\citealt{2017arXiv170202600H, 2017ApJ...841...24S}) calibrates the biased measurement of galaxy ellipticities on the image stamps. Each stamp is defined to be an image cutout with one galaxy. \textsc{metacalibration} creates one unsheared and four artificially-sheared galaxy stamps. Each is deconvolved with the input PSF, left unsheared or sheared with an additional gravitational shear of $\delta g=\pm 0.01$ in the two basis directions ($g_1$, $g_2$) and re-convolved with a new slightly larger isotropic, Gaussian PSF. We have chosen to use a Gaussian PSF since it simplifies the following (de)convolution operations and appears to be sufficient at our current testing precision. It will need to be verified that this continues to hold at the limit of the Roman requirements. \textsc{metacalibration} works under the assumption that the input shear is small, otherwise the recovered shear would be biased without accounting for higher order terms. It is a balance to choose an input shear that results in higher S/N of the estimated shear bias  and unbiased shear measurement. Figure \ref{fig:metacal_shear_linear} shows the performance as a function of input shear leading to the choice of input shear $g=\pm 0.02$. 

After processing the images with \textsc{metacalibration} and measuring the shapes, we have 5 shape catalogs for the five values of shear applied by \textsc{metacalibration}. We can then compute the shear response from these catalogs and calculate the calibrated ensemble galaxy shears. The shear response can be understood as the change in objects' ellipticities with respect to the change in gravitational shear. Mathematically, the observed ellipticities, $\vb*{e}$ = ($e_{1}$, $e_{2}$), can be approximated using a Taylor expansion around the two-component shear, $\vb*{g}=0$. Higher terms here can be dropped assuming the gravitational shear is small: 
\begin{equation}
    \vb*{e} = \vb*{e}\rvert_{\vb*{g}=0} + \frac{\partial \vb*{e}}{\partial \vb*{g}}\bigg\rvert_{\vb*{g}=0}\vb*{g} + \ldots
\end{equation}
From this expression, the shear response matrix can be defined as 
\begin{equation}\label{eqn:response}
    \vb*{R_{\gamma}} \equiv \frac{\partial \vb*{e}}{\partial \vb*{g}}\bigg\rvert_{\vb*{g}=0} = 
    \begin{pmatrix}
        \partial e_{1}/\partial g_{1} & \partial e_{2}/\partial g_{1} \\ 
        \partial e_{1}/\partial g_{2} & \partial e_{2}/\partial g_{2}
    \end{pmatrix}. 
\end{equation}
The ensemble mean of measured ellipticities $\langle\vb*{e}\rangle$ can then be computed. If the object selection cuts (e.g., S/N) are imposed when selecting the sample of galaxies for shear measurement, $\langle R \rangle = \langle R_{\gamma} \rangle + \langle R_{s} \rangle$, where $\langle R_{s} \rangle$ refers to the selection response 
\begin{equation}\label{eqn:select_response}
    \vb*{R_s} \equiv \frac{\partial \langle \vb*{e^{noshear}} \rangle}{\partial \vb*{g}}\bigg\rvert_{\vb*{g}=0}
\end{equation}
and corrects for selection bias. In this paper, all non-trivial object selection is done prior to image simulation, which prevents shear selection biases that arise due to object selection cuts imposed on the images; hence, $\langle R \rangle = \langle R_{\gamma} \rangle$ is a sufficient approximation.

Since we expect galaxies to be randomly oriented such that $\langle\vb*{e}\rangle\rvert_{\vb*{g}=0}=0$, we have a relationship between the measured ellipticities and the input shear in the simulations, 
\begin{equation}
    \langle\vb*{e}\rangle \approx \langle \vb*{R}\rangle\langle\vb*{g} \rangle. 
\end{equation}

In practice, we compute the mean shear response from the four sheared versions of the \textsc{metacalibration} catalogs using the finite difference method. 
\begin{equation}
    \langle\vb*{R_{ij}}\rangle = 
    \frac{\langle e^{+}_{i} - e^{-}_{i} \rangle}{2\delta g_{j}}, 
\end{equation}
where $e^{+}_{i}$ and $e^{-}_{i}$ are the $i$-th component of the galaxy ellipticity where positive ($+$) or negative ($-$) shear is applied in $i$-th direction. 

The mean shear response is then used to get the \textsc{metacalibration}-calibrated shears from the unsheared catalog. For individual objects, using Eqn.~\eqref{eqn:linear}, we compare the input shear and the calibrated shapes to calculate multiplicative ($\vb*{m}$) and additive bias ($\vb*{c}$). We note that these shear biases are computed using only the diagonal term of the shear response in Eqn.~\eqref{eqn:response}.

\section{Simulations}
\label{sec:sims}
The base of our image simulations of the \emph{Roman} Space Telescope is the \emph{Roman} simulation suite\footnote{\url{ https://github.com/matroxel/roman_imsim}} developed by T21, which renders realistic galaxy and star images on 18 Sensor-Chip Assemblies (SCAs) of a 2.5$\times$2.5 \textrm{deg}$^{2}$ patch of the sky following the observing strategy for the 5-year reference mission and Cycle 7 instrument specifications.\footnote{\url{https://roman.gsfc.nasa.gov/science/Roman_Reference_Information.html}} We begin by creating a truth catalog using the simulated galaxy distribution from the Buzzard simulation (\citealt{2019arXiv190102401D}), a photometric galaxy catalog sampled from the Cosmic Assembly Near-infrared Deep Extragalactic Legacy Survey (CANDELS; \citealt{2011ApJS..197...35G, 2011ApJS..197...36K, 2019ApJ...877..117H}), and a Milky Way simulation (Galaxia; \citealt{2011ApJ...730....3S}) for star positions and magnitudes. Then, the following properties are assigned to each galaxy:
\begin{itemize}
    \item positions (RA, Dec),
    \item flux within each \emph{Roman} filter (F184/H158/J129/Y106),
    \item intrinsic galaxy shapes and random orientations,
    \item flux ratios of de Vaucouleurs bulge, exponential disk, and star-forming knots, and
    \item artificial gravitational lensing shears.
\end{itemize} 
Positions are drawn from the galaxy density in the Buzzard simulation, while other properties are drawn randomly from realistic distributions, with intrinsic object properties following a distribution based on CANDELS data.
Within four identical realizations of the simulation, we use four sets of gravitational shears ($e_{1}$, $e_{2}$)=\{(+0.02, 0.00), (-0.02, 0.00), (0.00, +0.02), (0.00, -0.02)\}. This approach helps us to reduce shape and measurement noise when taking the difference in recovered shapes to compute the multiplicative bias (\citealt{2019A&A...621A...2P}).

The next step of the process is to create postage stamps and SCA images using \texttt{GalSim}. In this stage, the point-spread function (PSF) is convolved with the intrinsic galaxy light profile. Here, the \emph{Roman} PSF is rendered using the \texttt{galsim.roman} module, which has implemented \emph{Roman}-specific instrument properties such as the PSF and World Coordinate System (WCS) for a given telescope pointing, rotation angle, and SCA. The simulated detector effects were then added to the images in the following order: reciprocity failure, quantization, dark current, persistence, non-linearity, interpixel capacitance, read noise, and electron-to-ADU conversion (\citealt{2020JATIS...6d6001M}). These simulations have an achromatic PSF implementation, due to it being computationally infeasible to do a large chromatic simulation when the work began. For this analysis, we use isolated object cutouts for each object, so that effects related to blending can be ignored in this first study of \textsc{metacalibration} for \textit{Roman}.

After we generate the object stamps across all the SCAs and pointings, we create Multi-Epoch Data Structure (MEDS\footnote{Version 0.9.8 was used. \url{https://github.com/esheldon/meds}}; e.g., \citealt{2016MNRAS.460.2245J}) files in which each unique object dictionary contains information of all the exposures in which it appears in. These MEDS files are partitioned in a given Hierarchical Equal Area isoLatitude Pixelisation (\textsc{HEALPix}) of $n_{\text{side}}=512$. They contain all objects that are located in that region of the sky partitioned according \textsc{HEALPixel}\footnote{\url{https://healpix.jpl.nasa.gov/}} (\citealt{2005ApJ...622..759G, Zonca2019}).

Once the objects are sorted in MEDS files, we pass these multiple exposures with the corresponding PSFs to \texttt{ngmix}\footnote{Version 1.3.6 was used. \url{https://github.com/esheldon/ngmix}} to fit the galaxy shapes with the Gaussian mixture fitting method (\citealt{2014MNRAS.444L..25S}). This shape measurement process produces shape catalogs from which the shear response is calculated with \textsc{metacalibration}, and shear calibration bias can finally be computed.

\begin{figure}
	\includegraphics[width=\columnwidth]{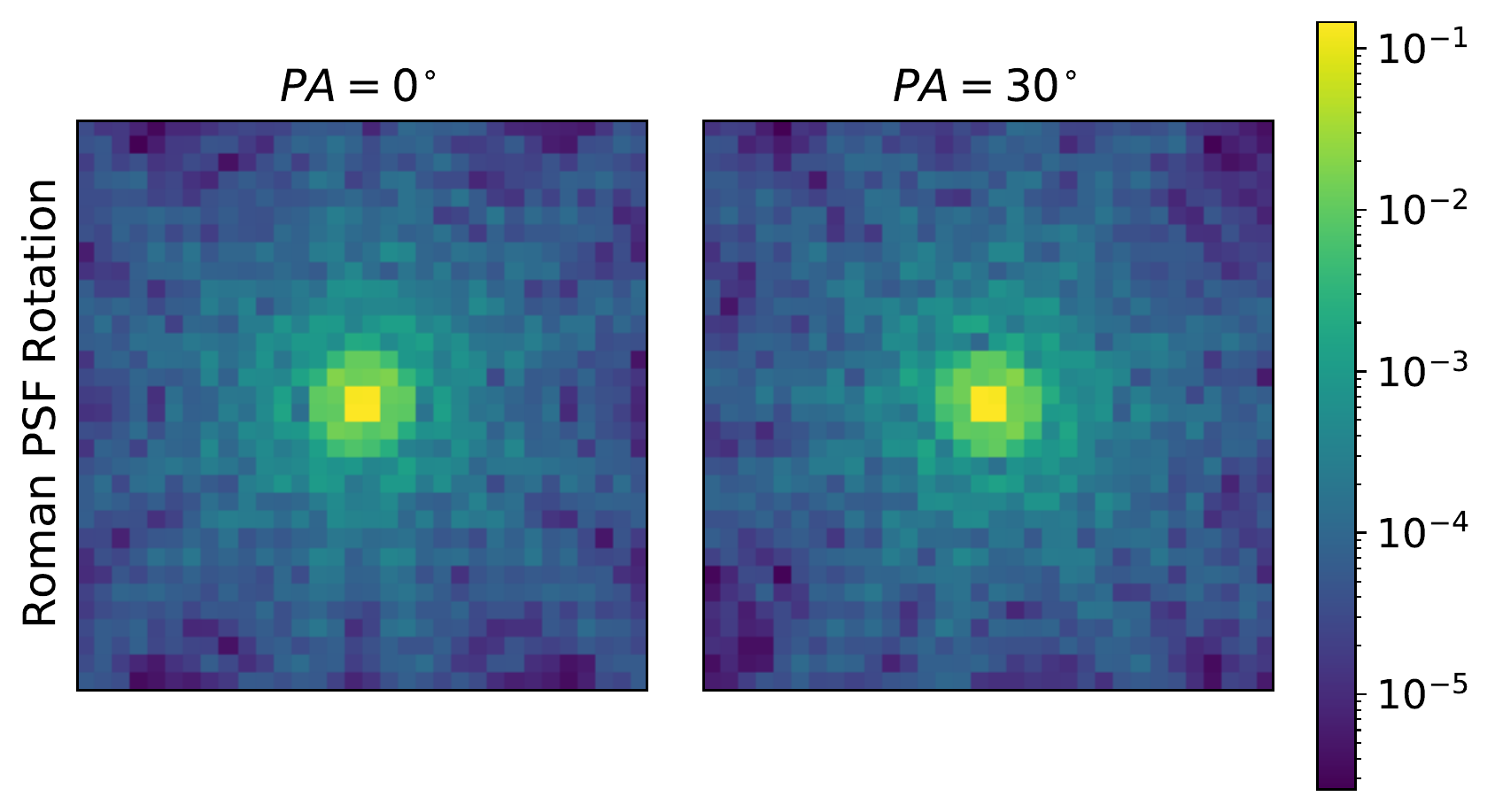}
    \caption{The rotation of the Roman PSF produced by the \texttt{galsim.roman} module. The position angle (PA) which determines the rotation of the focal plane is $0^{\circ}$ (\textbf{left}) and $30^{\circ}$ (\textbf{right}) clockwise. These are drawn for SCA=1 and H158 bandpass at a native pixel scale. The rotation of the PSF is particularly important when an object has multiple exposures. As more exposures are rotated relative to one another on the sky, the average impact of the PSF will be rounder. This will translate to a substantially less-elliptical coadd PSF.}
    \label{fig:psfrot}
\end{figure}

\begin{figure*}
	\includegraphics[width=\textwidth]{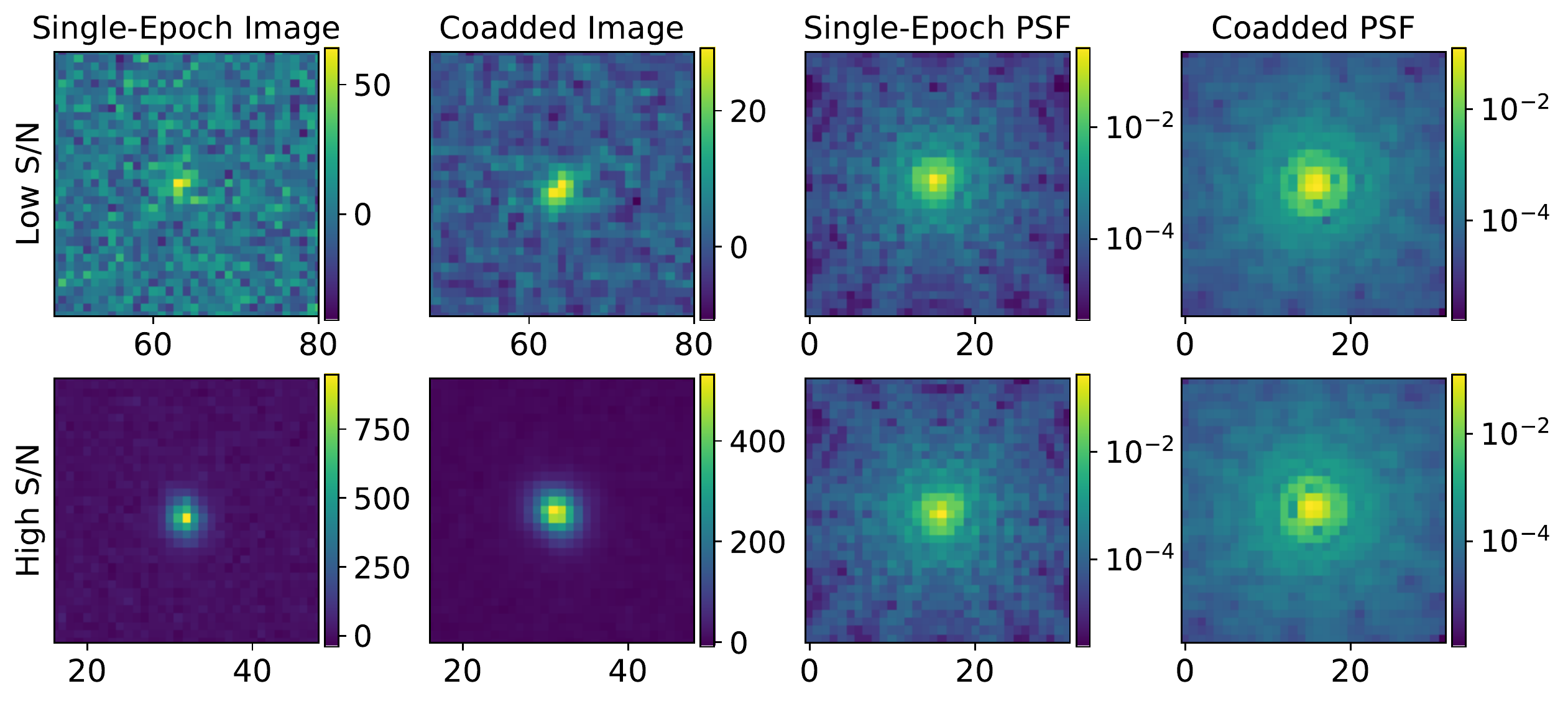}
    \caption{\textbf{Top Row}: the first observation of the single-epoch images and its coadded image, and the corresponding single-epoch and coadded oversampled PSF images for a galaxy with low signal-to-noise ratio (S/N=20 in a single-epoch image). \textbf{Bottom Row}: the same as above for a galaxy with high signal-to-noise ratio (S/N=390 in a single-epoch image). The H158 bandpass was used to represent images here. The single-epoch PSFs are almost identical, since there is no visually different feature for modeling them in different dithers and SCAs. There are two features to note here. One is how coaddition helps for low S/N objects, and the other feature is that the coadded PSFs look more isotropic than the single-epoch PSF due to different numbers of exposures.}
    \label{fig:singlecoadd}
\end{figure*}

\subsection{Updates to Simulation Capabilities}
\label{subsec:updates}
In order to accomplish the science goals, and build and test weak lensing calibration pipelines for \emph{Roman}, we have continued to update the realism of our image simulations. We have implemented and made updates to the following parts of the simulation framework to be able to better test shear calibration using \textsc{metacalibration}. 
\begin{itemize}
    \setlength\itemsep{1em}
    \item \textbf{Saturation Cuts}:
    We have implemented a pixel saturation limit of 100,000 electrons. This value is an expected pixel saturation level for \emph{Roman} detectors, but in the next generation of the simulations we will use a more accurate pixel saturation limit measured directly from the flight detectors. 
    
    \item \textbf{Rotation of the \emph{Roman} PSF on the sky}: In the study by T21, the rotations of the \emph{Roman} PSF with respect to the telescope rotation were not properly applied by the \texttt{galsim.roman} module. Since a non-rotating PSF produces an artificial preferred direction, which can translate to galaxy shapes through errors in the process of deconvolution, properly accounting for the averaging of the PSF orientation across exposures due to the survey dithering strategy is essential for a realistic shear calibration estimate. There has been an update in the \texttt{galsim.roman} module to correctly rotate the PSF given the WCS of the SCA in a given telescope pointing. Figure \ref{fig:psfrot} shows an example of how the PSF for one SCA changes with the rotation angle of the telescope.
    
    \item \textbf{Single-band and multi-band Coadds}: 
    We have used the postage stamp coadds (\texttt{psc}) algorithm\footnote{\url{https://github.com/esheldon/psc}} to create coadditions of single-epoch postage stamps of individual objects. Coadds are usually performed to enhance the signal-to-noise ratio (S/N) of an image and mitigate the impact of measurement noise, while allowing for better sampling of the image if the overlapping images are dithered. However, it is challenging to create a robust, continuous coadd PSF model over an entire image. This is mitigated if we instead construct small, local coadds (at the size of an object postage stamp cutout in this work), which is a method also utilized in \textsc{metadetection}. If we take advantage of fitting an object in the coadd instead of fitting multiple images of each object in each filter, we can save a factor of approximately six in computing time, since there are on average six exposures at any point on the sky in the reference survey. Additionally, coadding is one way to reduce the effect of the undersampling of the images since the coadd can be better sampled due to dithering. For these reasons, we explore utilizing this coadding scheme for the \emph{Roman} galaxies. In future work, a coaddition algorithm could be used such as \textsc{imcom\footnote{\url{https://github.com/barnabytprowe/imcom}} (\citealt{2011ApJ...741...46R})} that explicitly addresses the undersampling in a way that \texttt{psc} does not. The examples of the low and high S/N galaxy and PSF images for single-epoch and coadd are shown in Figure \ref{fig:singlecoadd}. The detailed overview of the coadd process in \texttt{psc} is described in in Sec.~\ref{subsec:psc}.
    
    \item \textbf{\textsc{metacalibration} in \texttt{ngmix}}: We implemented the \textsc{metacalibration}\footnote{Version 1.3.6 was used. \url{https://github.com/esheldon/ngmix/wiki/Metacalibration}} bootstrapper method, which is a wrapper class to run measurements in the \texttt{ngmix} package and was used to produce the weak lensing shape catalog in the DES Y3 analysis (\citealt{2020arXiv201103408G}). The \textsc{metacalibration} process can be performed on single-epoch or coadd images in each bandpass. With this calibration method, we hope to significantly reduce some of the shear calibration bias T21 measured in order to make comparisons of effects contributing to the bias more realistic and meaningful as we explore future \emph{Roman} pipeline development. The detailed overview of the \textsc{metacalibration} bootstrapper method can be found in Sec.\ref{sec:background}. 
\end{itemize}

\subsection{Simulated Weak Lensing Sample}
The way galaxies are simulated in this paper resembles the one in T21. We briefly overview how the galaxy properties are assigned from the CANDELS catalog and the galaxy distributions are provided using the Buzzard simulation. 

Among several galaxy properties, the morphological distributions are uniformly random (e.g., bulge/disk ratio), but the random intrinsic ellipticity distribution is appropriate to that expected from the data (e.g., the right ellipticity variance). The photometric, size, and redshift properties are all drawn consistently from the \emph{Roman} weak lensing sample predictions based on the CANDELS measurements (\citealt{2019ApJ...877..117H}). Since this CANDELS-based catalog was designed to be the representative of the ultimate \emph{Roman} weak lensing source sample, both the association of magnitude, size, and redshift for each individual object and the total distribution overall for our galaxies is consistent with what is expected for the \emph{Roman} measurements. It is also worth mentioning that our input catalog is actually the distribution of a noisy selection that has cuts in size and signal-to-noise (S/N>18) representative of the nominal selection of the \emph{Roman} weak lensing sample (\citealt{2018arXiv180403628D}). 

As mentioned in the beginning of this section, we use the Buzzard simulation to distribute the galaxies. In our simulations, only the projected distribution is selected randomly to enable a realistic projected clustering signal. The Buzzard has an upper redshift limit of 2.5. While we simulate the full Buzzard images, we isolate object stamps and only the isolated object stamps are added to the full image which are eventually used in the shape measurement for this work. 

This pre-selection of the galaxies ends up with the distribution that is close to the nominal weak lensing sample that we did not need to do a selection post-simulation.

\subsection{Simulation Runtime, Data Products, and Data Access}
The simulations here match those produced in T21, but with different true shear values and the updates mentioned in Sec. \ref{subsec:updates}, and so information about processing and data volume prior to the processing additions described in this paper match that provided in App.~D of T21.
The runtime of the shape measurement per MEDS region in the simulations and data volume on disk of the full shape catalogs (5 \textsc{metacalibration} catalogs for one of four shear sets) is summarized in Table \ref{tab:data} for each of the different simulation runs. The simulations are run on the Duke Compute Cluster\footnote{\url{https://oit-rc.pages.oit.duke.edu/rcsupportdocs/dcc/}}, with solid-state disks for the simulation I/O. The time it takes to process one MEDS file is shown in the table. The simulation is composed of a total of 480 MEDS files, for a total run time of about 2381 CPU hours for the multi-band coadd measurement.

\begin{table}
    \centering
    \begin{tabular}[width=\columnwidth]{l|c|c|c}
    \hline
    Measurement type &  CPU runtime (hours) & Catalog disk size  \\
    \hline 
    single-band multi-epoch  & 3.7 & 3.9 GB \\
    single-band coadd  & 2.9 & 3.8 GB \\
    multi-band multi-epoch  & 4.8 & 3.8 GB \\
    multi-band coadd  & 5.0 & 3.8 GB \\
    \hline
    \end{tabular}
    \caption{The runtime for a single CPU and disk size of the shape catalogs for different simulation runs. The runtime for multi-band measurements is slower due to having to copy, uncompress, and read from the three MEDS files (each 1GB compressed) for each bandpass. Runtime for coadd measurements also includes the postage-stamp coadding, which dominates the runtime over the metacal process. For comparison, the image generation of one SCA takes 1.8 CPU hours.}
    \label{tab:data}
\end{table}

\subsection{Validations with Simple Simulations}
\label{subsec:simplesim}
In addition to the main, realistic simulation of the \emph{Roman} reference HLIS survey, we also produce several sets of much faster, simple simulations  to verify that there are no obvious fundamental sources of systematic biases using the \textsc{metacalibration} method on undersampled space-based images. These simple simulations thus play a role in ruling out the sources of systematic biases that may a priori  potentially be a problem in the use of \textsc{metacalibration} with space-based images. 

The fixed parameters for the simulations are listed in Table \ref{tab:params}. We fix the original pixel scale of the \emph{Roman}, bandpass and input galaxy size. With these parameters, we simulated images with varying galaxy model profile, PSF model profile, artificial shear, and background noise level, without any complications in the images and pipelines such as detector effects and coaddition. We first chose simple light profiles for galaxies and PSF to be Gaussian (with PSF half-light radius of 0.089 arcsec, which matches the expected average PSF size for H158 of 0.178 arcsec from T21), and verified with \textit{Basic-0.02} simulation that our input gravitational shear ($\abs{g}$=0.02) is unbiased in the \textsc{metacalibration} framework and achieves the required multiplicative bias for \emph{Roman}. Figure \ref{fig:metacal_shear_linear} shows that our choice of applied shear is below the shear requirement; therefore, it will not be the contributor to the bias in the other simulation variants or the full simulation. We also tested the same setup with different input shears (\textit{Basic-0.05} and \textit{Basic-0.1}) to test where the weak shear approximation breaks down. The values of inferred bias are shown in Table \ref{tab:simple_sim_result}. 

Next, we tested the shape measurement pipeline by doubling the background noise with \textit{Doublesky-0.02}. This test should validate that the \textsc{metacalibration} process can tolerate the induced correlation of Poisson noise in the object profiles during the \textsc{metacalibration} shearing process, since it is not currently symmetrized in the process as the background noise field is (\citealt{2017ApJ...841...24S}). We find that this result is also consistent with zero, so we can confirm that the treatment of correlated Poisson noise does not trigger any bias at the level we can probe with the current simulations. 

In the final row in Table \ref{tab:simple_sim_result} (\textit{RomanPSF-0.02}), we finally validate that the use of the complex \emph{Roman} PSF instead of a Gaussian PSF does not bias the shape recovery even though it is undersampled. We note that these tests were performed before updating the PSF rotation, and so they do not benefit from rotationally-induced isotropy of the effective PSF. Since none of these obvious potential issues using simple Gaussian galaxy profiles will cause bias in $m$ at the level we can probe in the more complex simulation, we can more easily interpret results of the more complicated simulation and analysis pipelines described in the following sections. However, there remains a potentially concerning non-zero additive bias in $c_2$, particularly when using the Roman PSF model that will need to be studied further. 

\begin{table}
    \centering
    \begin{tabular}[width=\columnwidth]{|p{3cm}||p{3cm}|p{3cm}|p{3cm}|}
    \hline
    Parameter & Value \\
    \hline
    Pixel scale & 0.11 arcsec/pixel\\
    Bandpass & H158 \\
    Galaxy half-light radius & 1.0 arcsec\\
    Stamp size & Multiples of 32 pixels each side\\
    \hline
    \end{tabular}
    \caption{List of the parameter choices for the simple simulations. For all the simple simulation runs, we used a Gaussian galaxy profile and the magnitude for each object is drawn randomly from the \emph{CANDELS} catalog. For the stamp size, the multiplying factor was chosen so that 99.5\% of the flux is in the stamp.}
    \label{tab:params}
\end{table}

\begin{table*}
	\centering
	\begin{tabular}[width=\textwidth]{ c|c|c|c|c|c|c|c|c|c } 
		\hline
		simulation variants & galaxy profile & PSF profile & shear & background noise level & $m_{1}\times10^{2}$ & $m_{2}\times10^{2}$ & $c_{1}\times10^{4}$ & $c_{2}\times10^{4}$\\
		\hline
		Basic-0.02 & Gaussian & Gaussian & 0.02 & 5714.36 e-/arcsec$^2$ & 0.01$\pm$0.10 & -0.02$\pm$0.10 & -0.02$\pm$0.14 & 1.06$\pm$0.14\\
		Basic-0.05 & Gaussian & Gaussian & 0.05 & 5714.36 e-/arcsec$^2$ & 0.23$\pm$0.07 & 0.22$\pm$0.07 & 0.05$\pm$0.33 & 1.08$\pm$0.33\\
		Basic-0.1 & Gaussian & Gaussian & 0.1 & 5714.36 e-/arcsec$^2$ & 0.99$\pm$0.03 & 0.99$\pm$0.03 & 0.13$\pm$0.33 & 0.88$\pm$0.33\\
		\hline
		Doublesky-0.02 & Gaussian & Gaussian & 0.02 & 11428.72 e-/arcsec$^2$ & -0.12$\pm$0.18 & 0.01$\pm$0.18 & 0.06$\pm$0.36 & 1.04$\pm$0.36\\
		\hline
		RomanPSF-0.02 (non-rotated) & Gaussian & Roman & 0.02 & 5714.36 e-/arcsec$^2$ & 0.13$\pm$0.10 & 0.18$\pm$0.10 & -0.02$\pm$0.19 & 5.33$\pm$0.19\\
		\hline
	\end{tabular}
	\caption{This table compares the shear calibration bias, both multiplicative and additive bias for different simple simulation runs. In each simulation except the first and last row, 5 million objects were simulated. We simulated 15 million objects for the first and last row to obtain sufficiently small uncertainties.}
	\label{tab:simple_sim_result}
\end{table*}

\section{Coadd and shear calibration pipelines}
\label{sec:methods}
Our goal is to test whether the recovered shear with \textsc{metacalibration} is non-biased for \emph{Roman} and to understand the factors that might contribute to any non-negligible bias. In order to do so, we need to build measurement pipelines within the current simulation framework that produce a final calibrated shear measurement. In this section, we present in detail how coadd images are produced with \texttt{psc} from the undersampled single-epoch images and how \textsc{metacalibration} is implemented to calibrate the measured shear.

\subsection{Postage Stamp Coadds}
\label{subsec:psc}
Coaddition is the process of summing information from multiple overlapping images. If the single-epoch images are dithered, a Nyquist-sampled image can be constructed out of multiple undersampled exposures of an object. While this process can also be beneficial in increasing the S/N value of an object and reducing the impact of pixel noise, several challenges need to be addressed for images taken with telescopes that can rotate. When rotation is introduced, this coadding process becomes more complex to interpolate the image to stack a pixel grid. While coaddition can introduce new challenges and potential biases due to the complexity of coadding the original PSF and its interpolation scheme, potential bias due to the undersampling of the original images can be mitigated by appropriately coadding dithered images. 

Among imaging surveys, coadding a small region of the sky is common, for example \textsc{swarp} (Bertin et al. 2002) or \textsc{drizzle} (\citealt{2002PASP..114..144F}). However, we decided to coadd the postage-stamp cutouts to simplify the treatment of the coadd PSF. It is also beneficial to use this method because we need coadds which can be directly injected into the shape measurement pipeline in memory rather than written as images to disk due to the number of cutouts we have to process. We specifically use the simple interpolation-based coadding method using \texttt{GalSim}, \texttt{psc} (postage stamp coadds), as our coadding process.
While this method does not explicitly account for undersampling of the images and will produce aliasing at some level, we find that this is not an important factor at our current precision of tests in this work. Future work will explore more principled methods like \textsc{imcom} (\citealt{2011ApJ...741...46R}) that explicitly account for the undersampling of space-based imaging like our Roman image simulations.

We reconstruct a better-sampled coadd image from multiple exposures in MEDS files. For each exposure of the object we render the \emph{Roman} PSF with a stamp size of 32 pixels at the galaxy centroid. This choice of stamp size ends up losing about 2-4\% of the total flux of the PSF. However, we have no evidence that this impacts the shape measurement. We modified the original \texttt{psc} code to improve performance for this \emph{Roman} study, and these modifications are explained below with the general coadding process in \texttt{psc}. 

The algorithm: 
\begin{enumerate}
    \setlength\itemsep{1em}
    \item Finds the WCS of the first exposure of the object.
    \item Translates the original WCS to a flat WCS (locally diagonal Jacobian representation), because it produces more stable results with \texttt{ngmix}.
    \item Creates the coadd stamp with 0.8 $\times$ original pixel scale. We scale the original pixel scale of the final coadd stamp to increase the image sampling. This final pixel scale was chosen to prevent the presence of visual artifacts in the structure of the coadd PSF image. 
    \item Creates an interpolated image with \texttt{GalSim}  using a \texttt{lanczos3} interpolant and sums them. 
    \item Creates a coadded noise image from the weight of the original images. 
\end{enumerate}
Figure \ref{fig:singlecoadd} shows an example of the simulated single-epoch and coadded images and PSFs for objects with low and high S/N. Note that the shape and struts pattern in the original PSF can be isotropized by coadding the rotationally dithered PSFs.

\begin{figure*}
	\includegraphics[width=\textwidth]{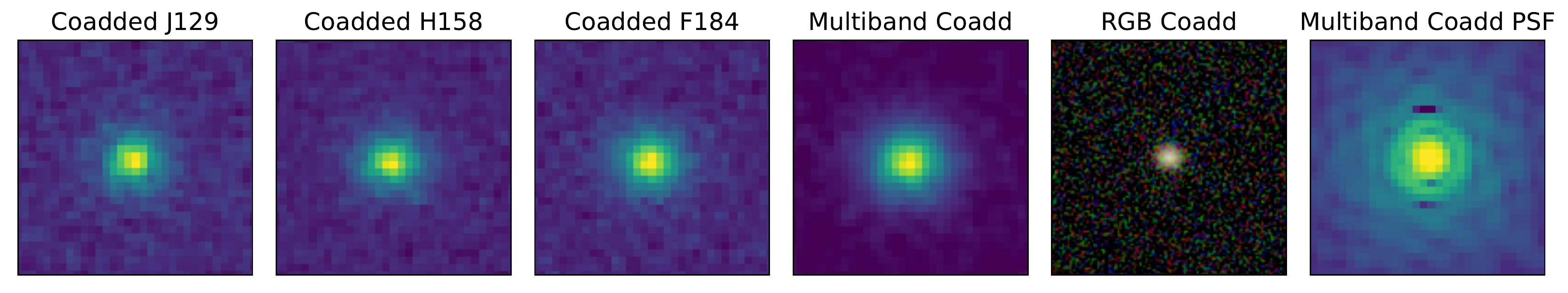}
    \caption{The left 3 panels show an example of the coadded galaxy images for each filter. The S/N for each filter is S/N=90, 68, 108, respectively. The 4th and 5th panels show the multi-band coadd images of the same object; each with linear scale and RGB scale. The S/N of the coadd image is S/N=110. Finally, the right-most panel shows the coadded multi-band PSF in a log scale. Some pixels with negative values were produced in the process of making the multi-band coadd PSF due to limitations in the fast interpolant functions used, but this has no impact on the results. }
    \label{fig:single_to_coadd_rgb}
\end{figure*}

Figure \ref{fig:single_to_coadd_rgb} shows the coadd products in different bandpasses for a relatively high S/N object.

\subsection{Shape Calibration \& Measurement with \texttt{ngmix}}
\label{subsec:mcal}
Once we construct the individual galaxy stamp coadds from the MEDS files, we recover the calibrated shear signal using the \textsc{metacalibration} process (\citealt{2017arXiv170202600H, 2017ApJ...841...24S}). In this work, the default interpolation kernel (\texttt{lanczos15}) was used to create the metacalibrated images, and the numerical tolerances used are also the default \texttt{GalSim} settings. The bootstrapper method used in \texttt{ngmix} wraps the \textsc{metacalibration} and the shape fit processes. The priors used for the fit are listed in Table \ref{tab:priors}.

The deconvolution process in  \textsc{metacalibration} was performed using the coadded PSF that is still undersampled. Since having a matching pixel scale in the galaxy and PSF image is a requirement for the version of the \texttt{ngmix} pipeline we implemented, we were not able to oversample the simulated \emph{Roman} PSF to utilize a more accurate PSF model. Future versions of these measurements should include the ability to provide a better-sampled PSF image for the deconvolution in the measurement. 
We also note that for the reconvolution process in \textsc{metacalibration} we have attempted the measurement with the \texttt{fitgauss}\footnote{An isotropic Gaussian PSF model fitted to the original PSF.} and \texttt{gauss} PSF model to reconvolve the sheared galaxy images, and the measurement with \texttt{fitgauss} model was numerically unstable resulting in the inconsistency between $m_{1}$ and $m_{2}$. We, therefore, use the \texttt{gauss} model for the rest of the simulations.

\begin{table}
    \centering
    \begin{tabular}{l l l}
    \hline
    Parameter & &  Prior \\
    \hline
    Pixel centroid offset & Flat & 0 < $p_{x,y}$ < 0.11\\
    Shear & Gaussian &$\langle g\rangle = 0.0$, $\sigma_{|g|} = 0.3$\\
    Galaxy size & Flat &$10^{-5}$ < T (arcsec$^2$) < $10^{4}$\\
    Flux fraction of the bulge  & Gaussian &$\langle f\rangle = 0.5$, $\sigma_{f} = 0.1$\\
    Total flux & Flat &$0$ < F (ADU) < $10^{6}$\\
    \hline
    \end{tabular}
    \caption{List of prior values and distributions used for the Gaussian mixture fit.}
    \label{tab:priors}
\end{table}

The covariances are computed using the bootstrap estimate of standard error. From the observed ellipticity catalogs, we randomly choose $n$ samples with replacement, where $n$ is the length of the data set, and compute the distributions of multiplicative ($f^{m}_{i}$) and additive ($f^{c}_{i}$) bias for $i=1,2,3,...,N$ ($N$ is the number of times the resampling is carried out) in the same way as above. The distributions are then used to compute the error estimate,

\begin{equation}
    \sigma_{N,f} = \sqrt{\frac{1}{N-1} \sum_{i=1}^{N}(f_{i}-\bar{f_{i}})^{2}}, 
\end{equation}
where we use sample $N$=200.

\section{Results}
\label{sec:results}
In this section, we present the properties of the simulated data products and the shape measurement results from various simulation variants. We divide our shape measurement into two categories; single-band and multi-band. For single-band measurements, in each filter, we measured the shapes from the original single exposures by jointly fitting them: we call this single-band multi-epoch measurement. We also measured shapes with the single-band coadd in each filter: we call this single-band coadd measurement. For the mutli-band multi-epoch measurement, we matched the objects between each filter and measured the shapes with the joint-fit of single exposures across all the filters. The multi-band coadd measurement was jointly fit across the three coadds from each filter to recover the shapes. 

\subsection{Statistics of the Shape Catalog}
Galaxies that are simulated are pre-selected to meet the \emph{Roman} Weak Lensing selection used for the mission requirements (\citealt{2018arXiv180403628D, 2021MNRAS.501.2044T}). In total, the galaxy photometry catalog contains 907,170 galaxies and from this catalog we simulated galaxies on 18 SCAs across 198 pointings for F184, 227 pointings for H158, and 238 pointings for J129 bandpass. We did not make any selection cuts on the simulated catalogs based on measured properties, since all input objects are selected to pass requirements for weak lensing selection. In our samples, no de-blending was necessary since the shape measurement was performed on object stamps without neighboring objects. Blending issues related to the object detection is mentioned later in Sec. \ref{sec:discussion}. 

\begin{figure*}
    \centering
	\includegraphics[width=\textwidth]{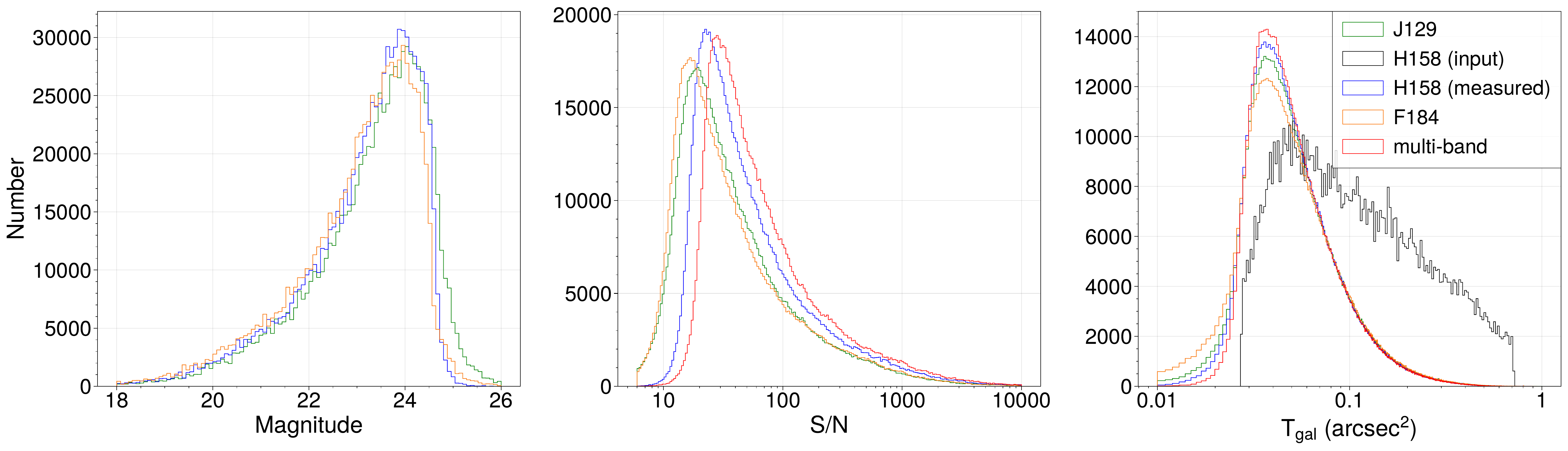}
    \caption{The histograms of input and measured galaxy properties in \texttt{ngmix} for single-band coadd and multi-band coadd measurements. \textbf{Left}: the input magnitudes of galaxies for all three filters we have used to measure shapes. The mean magnitude for J129, H158, F184 are respectively 23.3, 23.1, 23.0. \textbf{Middle}: the signal-to-noise ratio of the single-band and multi-band coadds. The median S/N for JHF and multi-band are respectively 29, 39, 27, and 50. \textbf{Right}: the input galaxy size for H158 approximated using a Gaussian profile with the half-light radius in the input H158 catalog, and the measured galaxy size $T$ for the single-band and the multi-band coadd in units of arcsec$^2$. The mean size for input and measured galaxies are 0.15 and 0.09 arcsec$^2$ for H158. The measured size is the size of the best fit Gaussian (maximum likelihood in this case) and it is not expected to match with the true size based on the half-light radius of the nominal (non-Gaussian) profiles.}
    \label{fig:ngmix_measured_properties}
\end{figure*}

In the end, we were able to measure the shapes of about 95\% of all the objects in the photometry catalog in each filter. This 5\% loss is mostly due to the fact that some objects are not saved in stamps due to being too large (i.e., require a stamp size of greater than 256 pixels). This selection is on the true size only, and so is not correlated with the shear. Some additional object shapes were not measured due to the Gaussian mixture fit in \texttt{ngmix} failing to converge. This also includes a very small number of objects that are rejected due to all cutouts being too near the SCA edges. The total number of the recovered objects for single-band measurements was, 861,407 for F184, 863,146 for H158, and 859,193 for J129. For the multi-band (H158+J129+F184) coadd measurements, we used objects that are measured successfully in all of the filters and the total number of the recovered objects is 851,821.
Figure \ref{fig:ngmix_measured_properties} shows the true magnitude and size, the measured signal-to-noise and size of galaxies in the \textsc{metacalibration} shape catalogs. There is a significant boost in signal-to-noise in the multi-band catalog vs the single-band catalogs. While the measured size ($T$) does not agree well with the input size derived from the half-light radius, this is a known issue due to fitting a Gaussian model to a galaxy with size defined as part of a much more complex profile.

\subsection{Shear Calibration Bias}
\label{subsec:shapes}
\begin{figure*}
	\includegraphics[width=\textwidth]{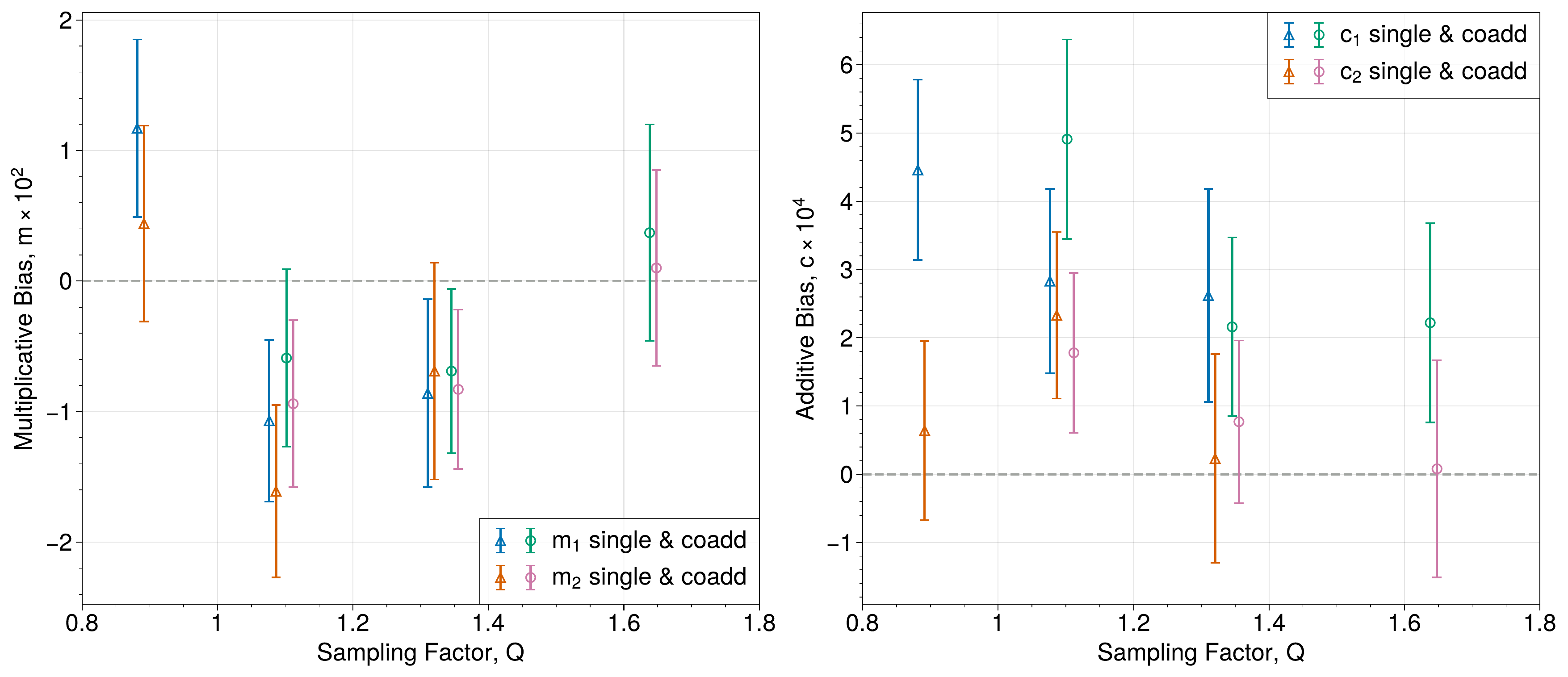}
    \caption{The multiplicative ($m \times 10^2$) (\textbf{left}) and additive ($c \times 10^4$) (\textbf{right}) bias of the single-band multi-epoch and coadd measurements. From the left to the right data points, they represent the data for each of the J129 multi-epoch, H158 multi-epoch, J129 coadd, F184 multi-epoch, H158 coadd and F184 coadd measurements. The data points for the $m_2$ and $c_2$ measurements are shifted by +0.01 for visual clarity. The non-calibrated results from the first-generation of the simulations by T21 were $m_1=(-7.56\pm0.19)\%$, $m_2=(-9.49\pm0.19)\%$ and $c_1=(1.20\pm0.17)\times10^{-3}$, $c_2=(-1.57\pm0.16)\times10^{-3}$. We can conclude that our calibration bias is an order of magnitude smaller than the previous versions of the simulations, and comparable or smaller than the bias-level seen in the Stage-III surveys.}
    \label{fig:final_result}
\end{figure*}

We estimate the levels of shear calibration bias associated with calibrating shapes with \textsc{metacalibration} in \emph{Roman} images as described in Sec. \ref{subsec:mcal}. The multiplicative and additive bias values listed in the text are the average of $m_{1}$ and $m_{2}$, and $c_{1}$ and $c_{2}$. The comprehensive results can be found in Table \ref{tab:bias_summary} and Fig. \ref{fig:final_result} shows the single-band single exposures and coadd results compared. These multiple simulation variants were run to achieve three milestones. One is to characterize how much the shear bias exists if we were to use the original, undersampled images, by recovering shears from the single exposures in each bandpass. Second is to verify if coadding can mitigate the effect of undersampling. Lastly, as the final assessment we investigate how combining the three bandpasses can result in better constraining the recovered shear.

For the single-band measurements, we can use the sampling factor defined in Equation \eqref{eqn:sampling} per bandpass to show the relationship between the shear bias and image sampling. The sampling factor is an indication of how undersampled the image is in each bandpass. The sampling factor for each single-band measurement can be found in Table \ref{tab:bias_summary}. This relationship between sampling factor and shear bias was previously explored in K21 using \emph{Euclid} simulation with different shape measurement algorithms, where they found that the estimated multiplicative bias has a relatively large dependence on the sampling factor. Their result can be extrapolated to our single-band multi-epoch results. The multiplicative bias of single-band multi-epoch measurements is consistent with zero within 2$\sigma$ in each filter, with mean uncertainty $\sigma_m$=0.72\%, except for $m_{2}=(-1.61\pm0.66)\%$ in H158, which is slightly larger. This level of bias is consistent with the finding from K21. 

When the coadded images for each filter are used, the recovered multiplicative shear bias is about half the multi-epoch cases, and is consistent with zero at the $\sim$1$\sigma$ level. The additive bias remains at a similar level within uncertainties. 

Finally, we discuss the results with multi-band measurements. We performed multi-band measurements with the three filters used in the previous measurements. One multi-band measurement used all of the single epoch images from all the filters and jointly fit the shape of the galaxy. The multiplicative and additive bias is, respectively, $m=(-0.76\pm0.43)$\% and $c=(2.56\pm0.79)\times10^{-4}$. This result generally agrees with the single-band multi-epoch measurements, and is consistent with zero bias at the $\sim$2$\sigma$ level. The other measurement is performed as a joint-fit across the coadd images for each filter. The multiplicative and additive bias is, respectively, $m=(-1.13\pm0.60)\%$ and $c=(2.38\pm1.24)\times10^{-4}$. These results are again consistent with zero at the 2$\sigma$ level, but have a large uncertainty relative to the Roman mission requirements. Scaling simply by volume, assuming a Gaussian error distribution, we would need a significantly larger simulation volume of about 2000 \textrm{deg}$^2$ in this complex simulation mode to draw firm conclusion about the performance of multiband measurements. This is comparable to the area of the reference Roman survey.

\begin{figure*}
    \centering
	\includegraphics[width=\textwidth]{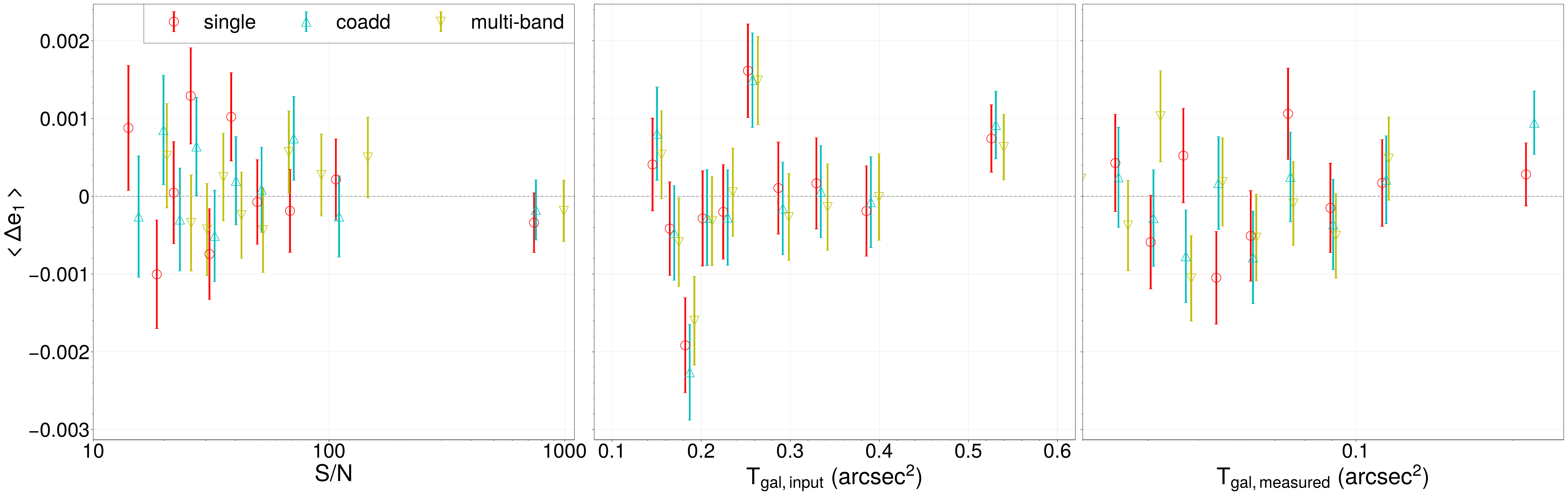}
    \caption{The mean shear residual ($\langle \Delta e_{1} \rangle$; $\langle e_{1,\text{measured}} \rangle - \langle e_{1,\text{expected}} \rangle$) as a function of various properties of galaxies for H158 single-band multi-epochs, coadds, and multi-band measurements. We are only showing the $e_1$ residual because $e_{2}$ showed similar behaviors in each test. We find results that are consistent with a null signal for signal-to-noise (left), true galaxy size (middle), and measured galaxy size (right). The mean shear is computed with the shear response and selection response in each bin to account for the selection bias. To make the figure more visible, blue points are shifted by -1.0 in S/N, -0.005 in $T_\text{gal,input}$, -0.001 in $T_\text{gal,measured}$, and yellow points are shifted by +1.0 in S/N, +0.005 in $T_\text{gal,input}$, +0.001 in $T_\text{gal,measured}$.}
    \label{fig:meanshear}
\end{figure*}

To further explore the shape catalogs, we performed a set of basic null tests and show representative results for $e_1$ for the H158 multi-epoch and coadd catalogs, and multi-band coadd catalog. The null tests should show that the mean shear residual ($\langle \Delta e_{1} \rangle$) is zero (flat as a function of galaxy properties) in the absence of any systematic biases. Figure \ref{fig:meanshear} shows the relationship between the mean shear and several input and measured properties of galaxies for all the measurement cases. We find no significant trends in mean shear vs. galaxy properties or vs PSF properties, which are not included in the figure.

\begin{table*}
	\centering
	\begin{tabular}[width=\textwidth]{ c|c|c|c|c|c } 
		\hline
		simulation variants & sampling factor ($Q$) & $m_{1}\times10^{2}$ & $m_{2}\times10^{2}$ & $c_{1}\times10^{4}$ & $c_{2}\times10^{4}$\\
		\hline
		single-band multi-epoch (J129) & 0.88 & 1.17$\pm$0.68 & 0.44$\pm$0.75 & 4.46$\pm$1.32 & 0.64$\pm$1.31\\
		single-band multi-epoch (H158) & 1.08 & -1.07$\pm$0.62 & -1.61$\pm$0.66 & 2.83$\pm$1.35 & 2.33$\pm$1.22\\
		single-band multi-epoch (F184) & 1.31 & -0.86$\pm$0.72 & -0.69$\pm$0.83 & 2.62$\pm$1.56 & 0.23$\pm$1.53\\
		\hline
		single-band coadd (J129) & 1.10 & -0.59$\pm$0.68 & -0.94$\pm$0.64 & 4.91$\pm$1.46 & 1.78$\pm$1.17\\
		single-band coadd (H158) & 1.35 & -0.69$\pm$0.63 & -0.83$\pm$0.61 & 2.16$\pm$1.31 & 0.77$\pm$1.19\\
		single-band coadd (F184) & 1.64 & 0.37$\pm$0.83 & 0.10$\pm$0.75 & 2.22$\pm$1.46 & 0.08$\pm$1.58\\
		\hline
		multi-band multi-epoch & N/A & -0.68$\pm$0.59 & -0.85$\pm$0.58 & 3.09$\pm$1.21 & 2.04$\pm$1.11 \\
		multi-band coadd & N/A & -1.02$\pm$0.62 & -1.24$\pm$0.57 & 2.37$\pm$1.26 & 2.38$\pm$1.22\\
		
		\hline
	\end{tabular}
	\caption{A comparison of the shear calibration bias, both multiplicative and additive bias, for different simulation runs.}
	\label{tab:bias_summary}
\end{table*}

\section{Future simulation needs and plan}
\label{sec:discussion}

We have investigated how \textsc{metacalibration} handles the complex \emph{Roman} PSF and undersampled images without accounting for the effect of blending. However, it has been found that galaxy blending at different redshifts could introduce a significant shear-dependent detection bias when calibrated with \textsc{metacalibration} (e.g., \citealt{2020ApJ...902..138S}, \citealt{2020arXiv201208567M}). In future studies, we will investigate the impact of blending in \emph{Roman} image simulations and implement an extended pipeline to correct shear-dependent blending/detection biases (e.g.,  \textsc{metadetection}) to explore more realistic shear calibration for the real survey.

Our team continues to increase the realism of the image simulations for future weak lensing calibration analyses. In the next generation of these simulations, we expect to include the following updates or upgrade these parts of the simulation. 
\begin{itemize}
    \setlength\itemsep{1em}
    \item \textbf{Simulation Volume}: Expand the simulated area by a factor of four to 20 \textrm{deg}$^{2}$, in order to obtain smaller error bars on the resulting shear calibration biases. Increasing the simulation volume has additional benefits beyond a more precise bias estimate, such as the ability to calibrate spatial dependence of shear and to measure two-point correlation function and propagate the biases into a cosmological analysis.
    
    \item \textbf{Pixel Masks}: Future versions of the simulations will take into account complex pixel masking of detector non-idealities, such as hot or dead pixels. This masking will be propagated through the analysis to provide realistic masking challenges. 
    
    \item \textbf{Near-Infrared Detector Effects}: Incorporate more realistic detector effects as measured in tests on the flight detectors (\citealt{2020JATIS...6d6001M}) and produce two versions of the simulation with and without them to determine the impact on the final cosmology result. The detector effects that are not simulated in this paper but will be incorporated are relative quantum efficiency, quantum yield, charge diffusion, brighter-fatter effect, burn-in, count rate non-linearity, classical non-linearity, vertical trailing pixel effect, bad pixels, gain, and the bias frame.
    
    \item \textbf{Image Coaddition}: As one of the scene coadd algorithms, we will implement a new coaddition strategy that coadds the whole scene instead of individual stamp cutouts using  \texttt{AstroDrizzle}\footnote{\url{https://github.com/spacetelescope/drizzlepac}}: A Python implementation of MultiDrizzle which was used for \emph{Hubble} Space Telescope (\citealt{2003hstc.conf..325B}). With \texttt{AstroDrizzle}, we are able to further mitigate the effect of undersampling in the \emph{Roman} images with smaller coadd pixel scales. We expect this to further reduce the measured shear bias based on its trend with sampling factor. As a more mathematically-principled method that directly accounts for the undersampling of the images, we can also evaluate the use of the \textsc{imcom}) algorithm (\citealt{2011ApJ...741...46R}) in the future.
    
    \item \textbf{Source Detection}: Implement a source detection and deblending methodology to allow more realistic tests of shear recovery.
    
    \item \textbf{Selection Cuts}: Simulate objects to below the detection threshold of the survey and incorporate standard quality assurance cuts (e.g., S/N) on the shape catalog. 
\end{itemize}

\section{Conclusions}\label{sec:conclusion}

In this paper, we explored the performance of the \textsc{metacalibration} shape measurement calibration algorithm and presented the first shear calibration result without the effect of blending on a realistic simulated version of the reference \emph{Roman} HLIS. Accurately characterizing how much shear estimates are biased using realistic simulated survey images is a requirement to successfully complete the weak lensing program of the \emph{Roman} HLIS. This early work allows us to develop such simulation resources and to benchmark current weak lensing methods to explore where effort is still needed in development for the \emph{Roman} mission. 

In a larger suite of simple simulations, where the galaxy and PSF profiles are Gaussian, we find that the shear calibration bias using \textsc{metacalibration} is consistent with zero at the $0.1$\% level, even though the images are undersampled. This finding persists even when using an accurate, complex \emph{Roman} PSF model. When exploring the performance of \textsc{metacalibration} in the current simulation framework where realistic complexities are incorporated, the runtime cost is significantly increased, and we can only constrain the shear calibration bias at the $0.6$\% level. In these simulations we include coadding the single-epoch cutout images and fitting the shapes from multiple bandpasses, and we find that shear estimates with \textsc{metacalibration} are unbiased at the $\sim 2\sigma$ level, which is similar to current-survey constraints. At the original image sampling level, the trend in bias vs sampling factor $Q$ is similar to what was previously found for \textsc{metacalibration} bias on undersampled images in the study by K21 using image simulations built for \emph{Euclid}, but at lower levels of bias. By coadding the cutout images and increasing the image sampling factor, we find that the residual bias is reduced by about a factor of two, but this is similar to the uncertainty. However, this is a promising result for using coadd-level shape measurement for \emph{Roman}, and we plan to further investigate the use of \textsc{metacalibration} with image coadds that can reconstruct Nyquist-sampled images in all bandpasses in future work.

These results are currently limited by the available simulation volume in the realistic simulated survey. To further constrain the shear calibration bias and validate that the \emph{Roman} mission will be able to achieve the weak lensing calibration required for the survey, we have outlined in Sec. \ref{sec:discussion} a variety of simulation improvements being planned in future versions of the \emph{Roman} simulations that are currently being developed, including an exploration of the impact of blending and object detection on the calibration. We will also need to simulate much larger volumes of data, which is a significant challenge in terms of computing resources and storage space. The next version of the realistic simulation will cover about a factor of four more of the sky. These first results with \textsc{metacalibration} in realistic simulated \emph{Roman} HLIS imaging demonstrate that the \textsc{metacalibration} shear calibration approach is a feasible strategy for the \emph{Roman} mission and undersampled imaging more generally. However, significant additional work to study shear calibration in the presence of blending and at significantly larger survey volumes is essential for reaching the precise requirements for the \emph{Roman} weak lensing mission. 

\section{Acknowledgements}

We thank Matthew Becker and Erin Sheldon for the useful discussions on the use of \texttt{ngmix}, and Arun Kannawadi for discussions about related \emph{Euclid} shape measurement work. This work was supported by NASA Grant 15-WFIRST15-0008 as part of the Roman Cosmology with the High-Latitude Survey Science Investigation Team (\url{https: //www.roman-hls-cosmology.space/}). This work used resources from the Duke Compute Cluster.

%%%%%%%%%%%%%%%%%%%%%%%%%%%%%%%%%%%%%%%%%%%%%%%%%%
\section*{Data Availability}

The new measurement catalogs produced in this work are available on reasonable request to the authors.



\bibliographystyle{mnras}
\bibliography{paper_v3} % if your bibtex file is called example.bib

%%%%%%%%%%%%%%%%% APPENDICES %%%%%%%%%%%%%%%%%%%%%

% \appendix

% Don't change these lines
\bsp	% typesetting comment
\label{lastpage}
\end{document}